\documentclass[sigconf]{acmart}

\usepackage{hyperref} 
\usepackage{amsmath, amssymb, amsthm, amsfonts} 
\usepackage{courier} 
\usepackage{float}
\usepackage{threeparttable} 
\usepackage{rotating} 
\usepackage{multirow} 
\usepackage{graphicx} 
\usepackage{lipsum}
\usepackage{booktabs} 
\usepackage{arydshln} 

\graphicspath{ {images/} } 

\usepackage{ulem}
\usepackage{wrapfig}

\begin{document}



\setcopyright{licensedusgovmixed}
\acmPrice{15.00}
\acmDOI{DOI URL HERE}
\acmISBN{ISBN HERE}
\acmConference[AutoSec '19]{ACM Workshop on Automotive Cybersecurity}{March 27, 2019}{Dallas, TX, USA}

\title[\Large{Towards a CAN IDS based on a neural-network data field predictor}]{Towards a CAN IDS based on a neural-network data field predictor}
\titlenote{  
  \tiny{This manuscript has been authored by UT-Battelle, LLC under Contract No. DE-AC05-00OR22725 with the U.S. Department of Energy.  The United States Government retains and the publisher, by accepting the article for publication, acknowledges that the United States Government retains a non-exclusive, paid-up, irrevocable, world-wide license to publish or reproduce the published form of this manuscript, or allow others to do so, for United States Government purposes.  The Department of Energy will provide public access to these results of federally sponsored research in accordance with the DOE Public Access Plan \url{http://energy.gov/downloads/doe-public-access-plan}.}
}



\author{Krzysztof Pawelec,\textsuperscript{1,2}  Robert A. Bridges,\textsuperscript{1} 
Frank L. Combs, \textsuperscript{1}}
\orcid{n/a, 0001-7962-6329, n/a}
\affiliation{%
  \institution{\textsuperscript{1}
  Oak Ridge National Laboratory, 
  \textsuperscript{2}
 The Pennsylvania State University}
}
\email{paweleckrzysztof1@gmail.com, bridgesra@ornl.gov, combsfl@ornl.gov }

\renewcommand{\shortauthors}{Pawelec, Bridges, \& Combs}









\begin{abstract}
Modern vehicles contain a few controller area networks (CANs), which allow scores of on-board electronic control units (ECUs) to communicate messages critical to vehicle functions and driver safety. 
CAN provide a lightweight and reliable broadcast protocol but is bereft of security features. 
As evidenced by many recent research works, CAN exploits are possible both remotely and with direct access, fueling a growing CAN intrusion detection system (IDS) body of research. 
A challenge for pioneering vehicle-agnostic IDSs is that passenger vehicles' CAN message encodings are proprietary, defined and held secret by original equipment manufacturers (OEMs). 
Targeting detection of next-generation attacks, in which messages are sent from the expected ECU at the expected time but with malicious content, researchers are now seeking to leverage ``CAN data models'', which predict future CAN message contents and use prediction error to identify anomalous, hopefully malicious CAN messages. 
Yet, current works model CAN signals post-translation, i.e., after applying OEM-donated or reverse-engineered translations from raw data. 
In this work, we present initial IDS results testing deep neural networks used to predict CAN data at the bit level, thereby providing IDS capabilities but avoiding reverse engineering proprietary encodings. 
Our results suggest the method is promising for continuous signals in CAN data, but struggles for discrete, e.g., binary, signals.  
\end{abstract}
%
%



\settopmatter{printfolios=true} 

\maketitle

\section{Introduction \& Background}
\label{sec:intro}
Modern vehicles are increasingly ``drive-by-wire'' meaning once-mechanical interfaces of subsystems have been replaced by communication of electronic control units (ECUs), or small computers orchestrating the subsystems. 
Rather than using dedicated connections for each  ECU pair,  a few controller area networks (CANs) allow broadcast communications of all ECUs. 
In particular, we focus on the high-speed (250Kbs-500Mbs) controller area network (CAN) bus, as it is used for much of modern vehicle communications.

\begin{figure}[ht]
\vspace{-.2cm}
\includegraphics[width=.48\textwidth]{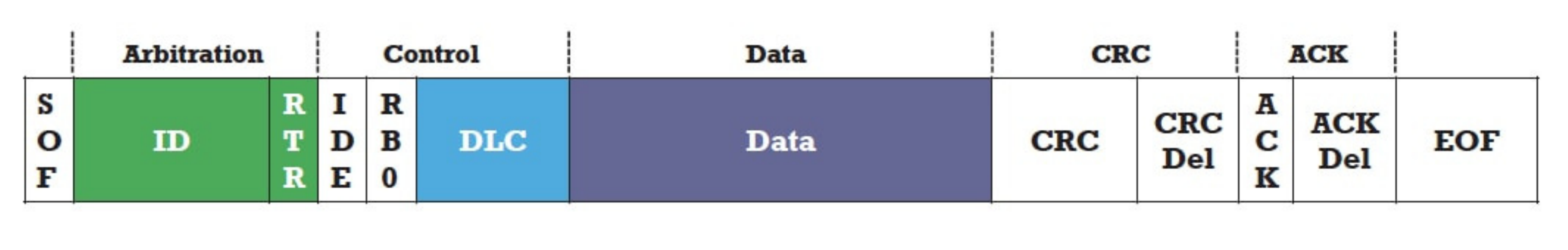}
\caption{
CAN 2.0 data frame depicted. Image from Cho \& Shin \cite{cho2016fingerprinting} of. There are two important fields, the Arbitration ID (AID) used for indexing and prioritizing frames and the data field containing up to 64 bits of message contents.
}
\label{fig:CANframe}
\end{figure}   

CAN 2.0 provides a standard protocol defining the physical and data link layers \cite{Bosch_GmbH_1991}. 
See Figure \ref{fig:CANframe} for the automotive CAN frame format. 
Each packet's information is contained in two fields, the Arbitration ID (AID)  
used for indexing and prioritizing frames and the data field containing up to 64 bits of message contents. 
The mapping of the data field's bits to the signals it encodes is a proprietary secret, defined by the original equipment manufacturers (OEMs, e.g., Ford, GM), and the encodings change depending on make, model, year, and even vehicle specifications.  
This poses an obstacle for producing \textit{vehicle-agnostic} solutions for  automotive CANs, in particular, defensive and offensive cyber security. 
See recent work of Verma et al. \cite{verma2018actt}, and Nolan et al. \cite{nolan2018unsupervised} on discovering the syntax and semantics of automotive CAN data.   

CAN is a reliable and lightweight protocol, but it has few security features, e.g., no encryption nor authentication, and has been proven to be exploitable with direct access \cite{hoppe2008security, checkoway2011comprehensive, moore2017modeling, miller2013adventures} or even remotely \cite{miller2015remote, woo2015practical}. 
The attack surface for in-vehicle CANs is growing as cars become increasingly exposed e.g. via USB, cellular, bluetooth and the advent of vehicle-to-vehicle and -infrastructure networking. 
Providing effective intrusion detection for automotive CANs is a burgeoning research topic \cite{tomlinson2018towards}.

\subsection{Related CAN IDS Works}
Initial automotive CAN IDS research has been rule-based \cite{muter2010structured, hoppe2008security}, which pushes security to OEMs, as rules are dependent on CAN encodings (model-specific) and may require knowledge of specific attacks. 
Multiple works \cite{moore2017modeling, gmiden2016intrusion, song2016intrusion} exploit message frequency anomalies for  vehicle-agnostic detection of message injection attacks. 
In response to the infamous Miller and Valesek remote Jeep hack \cite{miller2015remote}  (which used a masquerade attack in which one ECU sent malicious braking signals while the brake ECU was silenced),  
multiple efforts have proposed data-driven efforts for ECU identification to detect AIDs originating from the wrong transmitter \cite{cho2016fingerprinting, leeotids, choi2018identifying}.

The logical next-generation attack involves a reprogrammed ECU sending appropriate AIDs with appropriate timing, but with augmented, potentially malicious, data field contents. 
After-market ``chipping'' kits exhibit this capability by reprogramming ECUs, although in practice these are used for performance-tuning, not malicious purposes. 
Works are emerging that test supervised deep learners trained on specific attacks with labeled data \cite{kang2016intrusion, Loukas_Vuong_Heartfield_Sakellari_Yoon_Gan_2018}. 
We seek anomaly detection to avoid training towards a specific attack. 

Unsupervised CAN IDS research for detecting malicious message contents has begun modeling correlations inherent to the CAN data that may be broken by such attacks, admitting detection. 
Tyree et al. \cite{tyree2018exploiting} propose a manifold learning technique to identify relationships in CAN data that are broken during attacks that do not coordinate related signals. 
Their technique requires at least the ability to tokenize (partition) the up-to 64-bit CAN data fields into signal-sized messages but not fully translate the CAN data. 
The other three works seeking to exploit CAN data correlations require complete knowledge of the CAN signals:
Ganesan et al. \cite{2017-01-1654} learn correlation of value pairs (e.g., speed, accelerator pedal position) using both CAN  and sensor data to detect injection attacks.    
IDS research of Li \cite{li2016deep} and of Testud \cite{testud2017detecting} propose a three-step process to model CAN packets and detect unexpected packets: 
(1) reverse engineer or partner with an OEM to obtain many signals in the CAN data,
(2) train deep learning, neural network regressor(s) to predict the next signal value(s) from the history of observations, 
(3) use the error in predicted values from observed as an online anomaly detector. 

We present initial results for a CAN prediction model  without step (1). 
That is, previous work translated the 64-bit data field into the signals it encodes (requiring OEM knowledge or tedious reverse engineering) and built models of the signals. 
Rather, our approach models an AID's 64-bit data field. 
Hence, we commence prediction and detection (steps (2) and (3)) without requiring any translation of the CAN message bits to signals.

\subsection{Contributions}
Our long-term goal is to provide an after-market IDS for ideally all passenger vehicles. 
This means we cannot rely on OEM-defined CAN mappings. 
En route to this goal we adopt the neural network CAN prediction model; specifically, from a history of CAN data our regressors predict the next CAN data field, and we too use prediction error to detect anomalous messages. 
Unlike the previous two similar works \cite{li2016deep, testud2017detecting}, we do not translate CAN data fields to signals, as we do not have the OEM's proprietary mappings. 
Instead, we train a deep neural network for each  AID to predict its next 64-bit data field. 

The primary contribution of this work is presentation of initial results showing efficacy of the bit-level CAN models for attack detection. 
The benefit of this approach is straight-forward---it extends the general CAN modeling frameworks for anomaly detection (which relies critically on OEM-proprietary CAN mappings) to a vehicle-agnostic detector, as no CAN mappings are assumed. 
Although our focus is CAN IDS, CAN models can be used for other applications, e.g. CAN simulators.

\begin{wrapfigure}[18]{r}{.18\textwidth}
\vspace{-.3cm}
\includegraphics[width=.18\textwidth]{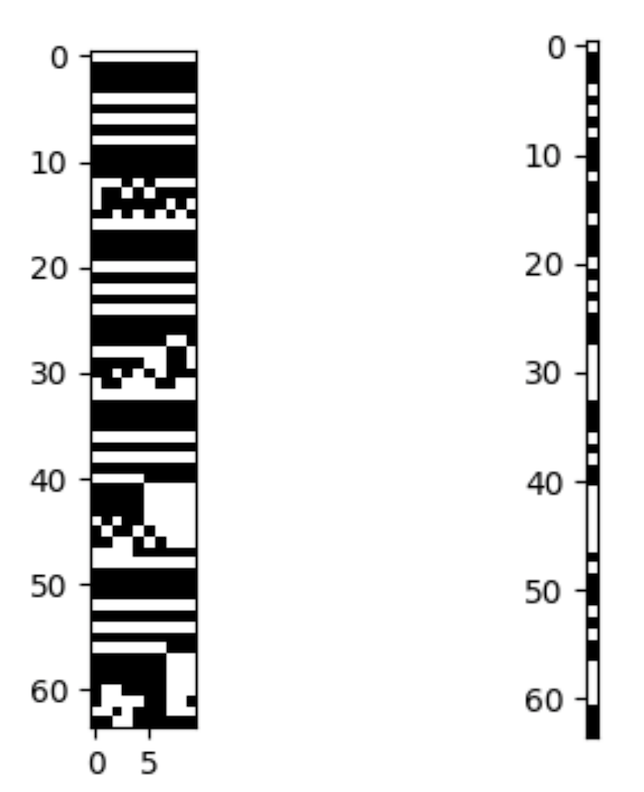}
\caption{Training data example, on the left, ten consecutive signals are labeled by 11\textsuperscript{th} (right).}
\label{fig:training-example}
\end{wrapfigure}
\section{CAN Prediction Model} 
The essential hypothesis of CAN prediction models is that there exists a dependency of future messages on recently passed or other concurrent messages. 
While our overall attack detector is unsupervised\textemdash that is, we do not require labeled attack and non-attack data\textemdash we exploit supervised learning to build a CAN prediction model. 
Specifically, we create labeled data by taking a fixed AID's most recently observed ten data fields  and try to predict next (11\textsuperscript{th}) one. 
Hence, we model each AID independently. 

Let $\mathbb{X}=\{x_i\}_{i=1}^{N}$ be the set of training examples and $\mathbb{Y} =\{y_i\}_{i=1}^{N}$ be the set of labels. 
Our training data is a tuple $(x_i, y_i)$ where $x_i \in \{0,1 \}^{10 \times 64}$ and $y_i \in \{0,1 \}^{64}$  as shown in Figure \ref{fig:training-example}.

\begin{figure}[b]
\centering
\vspace{-.25cm}
\includegraphics[width=.5\textwidth]{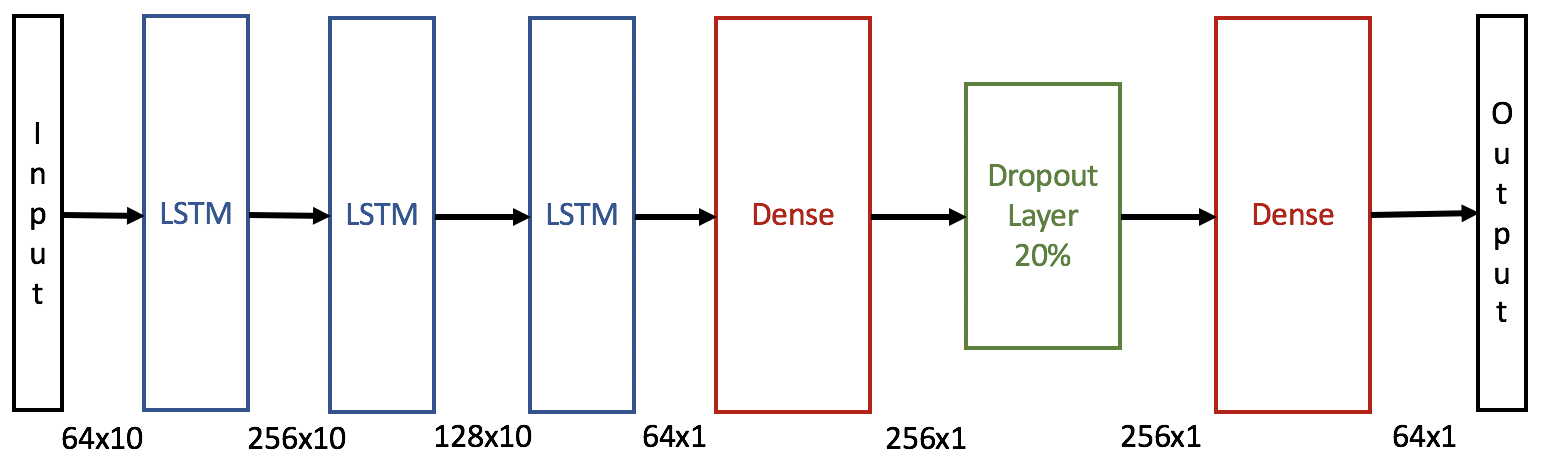}
\caption{Neural Network architecture diagram depicted. Dimensions of the vector passed between layers given. 
Batch size was set to 32. Dropout between dense layers set to randomly ignore 20\% of neurons in the first Dense layer.}
\label{fig:architecture}
\end{figure}

Recurrent neural networks (RNNs) model temporal/sequential dependence by including the previous prediction's hidden state as well as given  inputs into the current prediction \cite{rumelhart1986learning}. 
Long Short-Term Memory (LSTM) layers provide a particular architecture for portions of an RNN that seek to leverage dependence in modeling better than ``vanilla'' RNNs, as they are crafted to avoid vanishing gradient problems common in RNN  training \cite{Hochreiter_1997}. 
Hence, this statistical machinery is a natural choice for our model.

We build the model using Keras (\url{www.keras.io}), a Python deep learning module. 
The model consists of three LSTM layers, a dropout layer, and two dense layers. The last layer having 64 nodes (one per predicted bit of the next data field) as the output and softmax as an activation function.
Between the two dense layers, we include a dropout layer to prevent overfitting of our model. We set the layer's drop rate to 0.2 (i.e. 20\% of neurons in the first Dense layer are dropped during training). To train the model we used batch size of 32. Out of several tested architectures, where we varied number of layers and size of the hidden layers, this one showed best performance.
See Figure \ref{fig:architecture}.

For each desired AID, we use the described LSTM on ambient CAN data collected during normal driving conditions. 
We denote such a model $\mathcal{M} = \mathcal{M}_{(\mathbb{X},\mathbb{Y}, \text{AID})}$, where $\mathbb{X}$ is the set of training examples and $\mathbb{Y}$ is the set of labels for each example. 
For a given input vector $x_i \in \mathbb{X}$ (previous ten observed data fields), let $\mathcal{M}(x_{i}) = \hat{y}_i$ denote the predicted next 64-bit data field, $y_i \in \mathbb{Y}$.

To build an anomaly score from the AID's trained prediction model, we consider the error of each prediction, $e_i: = \|y_i -\hat{y}_i\|_2$. 
To account for model inaccuracies, compute the mean and variance of the observed prediction errors by using the model on the training set. 
Specifically, $\mu = \sum_{\mathbb{X}}e_i/N,$ and  $\sigma^2  =\sum_{\mathbb{X}}(e_i -\mu)^2/(N-1)$. 
Finally, we compute the Gaussian $z$-score of newly observed error $z_i = (e_i -\mu)/\sigma$ and use the one-sided p-value for our anomaly score, p-value$(z) = 1-\text{CDF}(z)$, where CDF is the Gaussian normal cumulative distribution function. 
Note that if the error is less than expected ($z < 0$ ) p-value$(z) > 0.5$ and p-value$(z) \to 1 $ as $z\to-\infty$. 
Similarly, if the error is greater than expected ($z > 0$) p-value$(z) < .5$, and p-value$(z) \to 0 $ as $z\to \infty$. 
Hence, a small p-values occurs if and only if the error is large relative to observations in training.


\section{Experiment}
\label{sec:experiment}
To respect space constraints, we present two indicative experiments, one model of an AID that seems to communicate continuous signals, another of an AID that seems to communicated discrete signals. 
Specifically, we believe the first AID communicates four two-byte messages giving the wheels' respective speeds and the second AID a binary indicator for if the vehicle is in reverse. 
For data collection we used the Vehicle Spy software, produced by Intrepid Control Systems, Inc. (\url{www.intrepidcs.com/products/software/vehicle-spy}) allowing passive monitoring of CAN data via the OBD-II port. 

\begin{figure}[b]
\includegraphics[width=.47\textwidth]{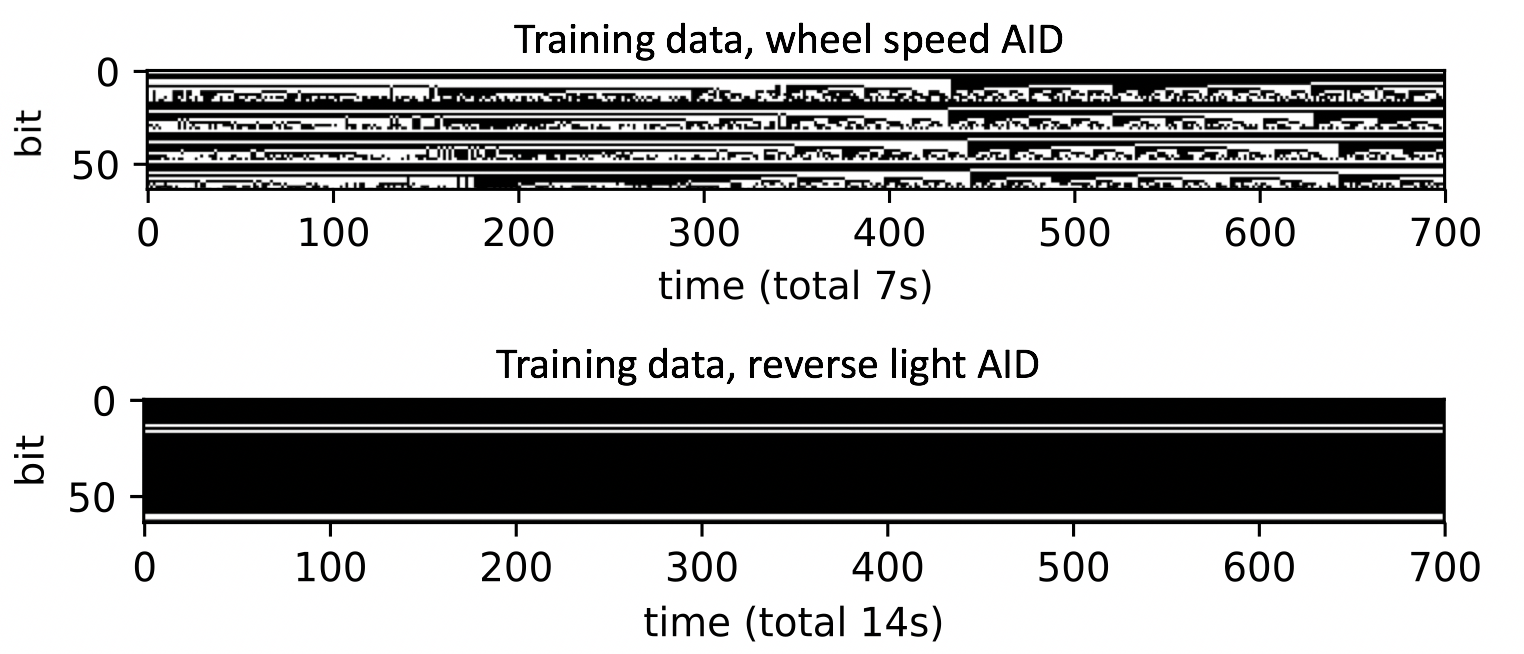}
\caption{Time snippet of training data for two AIDs.}
\label{fig:training-data}
\vspace{-.5cm}
\end{figure}

For training, we used a portion of CAN data recorded during ambient driving lasting 141 seconds. 
Figure \ref{fig:training-data} visualizes a snippet of the training data for each AID. 
Once the prediction models are trained for each AID, we must fit a Gaussian to the observed prediction errors using only the training examples. 
Hence, we apply the trained model to the training set and observe the prediction errors $\{e_i\}$, then  compute the mean,  $\mu$ and variance $\sigma^2$. 

To test the detector, we inject  CAN frames with each AID, separately,  to emulate attacks on the CAN.  
It is important to stress that the anomaly detector does not consider the frequency nor the timestamp of  CAN frame, only the sequence of data fields; 
hence, the high frequency injections emulate an ECU that is sending messages with false content. 
For each emulated attack (one per AID), we used an Arduino board for injecting CAN frames as well as the Vehicle Spy for recording CAN data,  both connected to the vehicle  via an OBD-II port.

\subsection{Wheel Speed AID}
\begin{figure}[h]
\vspace{-.5cm}
\includegraphics[scale=0.21]{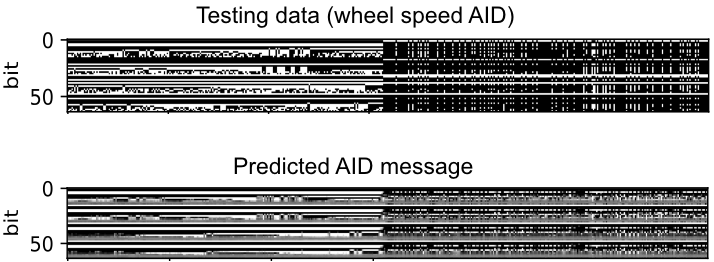}\\
\includegraphics[width = .4\textwidth]{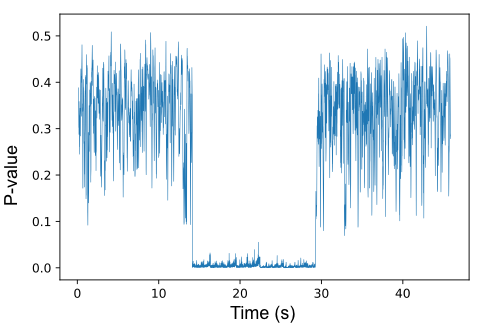}
\caption{(Top) Time snippet of wheel speed AID testing data non-attack period (the left half) and the attack period (the right half). (Middle) Plot depicts the model's prediction. (Bottom) P-value anomaly score depicted for wheel speed AID. Attack period: 14-29s. }
\label{fig:wheelspeed}
\label{fig:p-value-wheelspeed}
\vspace{-.3cm}
\end{figure}

The actual attack happened from 14s to 29s of the trip. During that time the ``attacker'' repeatedly injected the same AID with the same message in the 64-bit data field.
As can be seen in Figure \ref{fig:p-value-wheelspeed}, the p-value of the observed signals occurring between 14s to 29s is extremely low. 

\subsection{Reverse Lights AID}
\begin{figure}[h]
\vspace{-.4cm}
\includegraphics[scale=0.28]{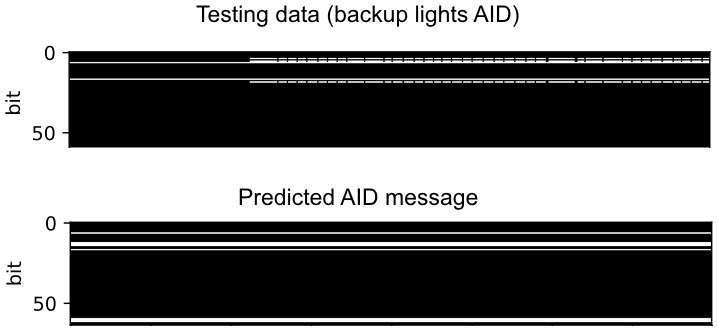}\\
\includegraphics[width = .4\textwidth]{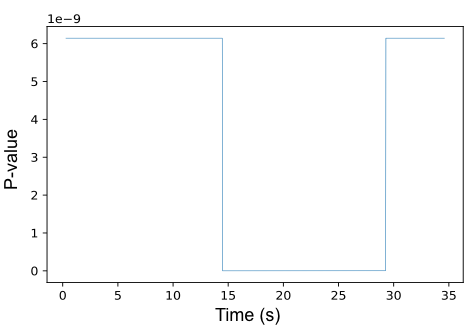}
\caption{(Top) Time snippet of wheel speed AID testing data. Non-attack period occurs for roughly the first quarter and the attack period in the other three quarters. Top plot depicts actual data. (Middle)Plot depicts the model's prediction.
(Bottom) P-value anomaly score depicted for reverse light AID. Attack period: 14.5-29s.}
\label{fig:reverse-lights}
\label{fig:p-value-backup-light}
\vspace{-.6cm}
\end{figure}

The actual attack happened from 14.5s until 29s of the capture. 
During that time the ``attacker'' repeatedly injected the same AID with the same message in the 64-bit data field. 
Referring to Figure \ref{fig:p-value-backup-light},  it is important to note that the p-value of the observed signals is extremely low throughout the test set. 
However, it hits actual 0 during the attack period. 

\subsection{Results Discussion}
Overall, we have a very strong difference in our anomaly score between attack and non-attack periods, but finding an a priori threshold seems problematic. 
We conjecture that current architecture is a better model for nearly continuous signals with many distinct 64-bit messages (as in Figure \ref{fig:wheelspeed}), that move in a a clear pattern (e.g., as speed increases, the $2^0$ place bit increases from 0 to 1, then the $2^1$ place bit increases from 0 to 1, ... ). 
The second AID communicating seemingly binary signals is, unsurprisingly, harder for the model to predict. 
Perhaps taking inputs from a variety of other AIDs may enhance prediction accuracy. 

\section{Conclusions \& Future Work} 
\label{sec:conclusion}
Recent approaches to build CAN IDSs train a ``CAN language model'', that is, a machine learning model that can accurately predict the next CAN message from previous or concurrent messages. 
Previous works have trained models on reverse engineered signals, requiring OEM-proprietary (secret) knowledge. 
In this paper we build a CAN model at the bit level, eliminating the need for CAN data translation, and present initial results in use for an IDS. 

To build the CAN model, we assumed a dependency between previous and future data fields within an AID of an automotive CAN, and train an LSTM recurrent neural network on ambient data for two AIDs. 
From both AID CAN models, we build an anomaly detector based on a relative predicted error of each CAN message. 
A very important feature of our method is that our neural network takes on raw 64-bit messages, and hence does not require extensive preprocessing, e.g., to reverse engineer proprietary CAN encodings. 
The technique works very well with AIDs that carry many distinct messages (roughly continuous messages) which change often over time. 
On the other hand, applying the same neural network architecture to an AID with seemingly binary signals (and therefore few distinct messages)
does not yield as convincing results.  
In particular, prediction error during the non-attack period during testing was very large relative to expectations from training (Fig. \ref{fig:p-value-backup-light}). 
 
For future work, we would like to refine the architecture of the neural network to more accurately predict non-malicious messages. 
Although outside of scope for this paper, we note that preliminary testing with alternate neural network configurations yielded less accurate results, but lends credence to future work aimed at optimizing the architecture for CAN modeling. 
Additionally,  construction of a model that handles more than just one AID at a time will presumably increase accuracy as  CANs communicate states of many different but physically related  subsystems.  
Finally, work is emerging to automatically discover encoded signals in the CAN data fields (e.g, \cite{verma2018actt, nolan2018unsupervised}); hence, the logical next step is to train the CAN models conditioned on information from these works. 

 
\section*{Acknowledgements}
\label{sec:acks} 
Authors thank S. Hollifield,  J. Laska, and M. Verma for fruitful discussions. 
Research sponsored by the Laboratory Directed Research and Development Program of Oak Ridge National Laboratory, managed by UT-Battelle, LLC, for the U. S. Department of Energy and the  National Science Foundation Math Science Graduate Internship (NSF-MSGI). 

\bibliographystyle{ACM-Reference-Format}
\bibliography{vsl} 

\end{document}